\newcommand*{\ATLASLATEXPATH}{}
\documentclass[PAPER, cernpreprint, texlive=2018, LANGEDIT=true,UKenglish]{\ATLASLATEXPATH atlasdoc}
\LEcontact{Richard Keeler, rkeeler@uvic.ca}
\usepackage[subcaption]{\ATLASLATEXPATH atlaspackage}
\usepackage{\ATLASLATEXPATH atlasbiblatex}
 
\usepackage{\ATLASLATEXPATH atlasphysics}
 
\graphicspath{{logos/}{figures/}}
 
\usepackage{ANA-HIGG-2018-42-PAPER-defs}

\AtlasTitle{A search for the $Z\gamma$ decay mode of the Higgs boson in $pp$ collisions at $\sqrt{s}$ = 13~TeV with the ATLAS detector}
 
\PreprintIdNumber{CERN-EP-2020-052}
\AtlasAbstract{
A search for the $Z\gamma$ decay of the Higgs boson, with $Z$ boson decays into pairs of electrons or muons is presented.
The analysis uses proton--proton collision data at $\sqrt{s}$ = 13~TeV corresponding to an
integrated luminosity of 139~fb$^{-1}$ recorded
by the ATLAS detector at the Large Hadron Collider.
The observed data are consistent with the expected background with a $p$-value of 1.3\%.
An upper limit at 95\% confidence level on the production
cross-section times the branching ratio for $pp\to H\to Z\gamma$ is set at
3.6~times the Standard Model prediction while 2.6~times is expected in the presence of the Standard Model Higgs boson.
The best-fit value for the signal yield normalised to the Standard Model prediction is $2.0^{+1.0}_{-0.9}$ where the statistical component of the uncertainty is dominant.
}
 
\author{The ATLAS Collaboration}
 
\AtlasRefCode{HIGG-2018-42}

\AtlasJournal{Phys.\ Lett.\ B.}
 
\AtlasJournalRef{Phys. Lett. B 809 (2020) 135754}
 
\AtlasDOI{10.1016/j.physletb.2020.135754}
 
\hypersetup{pdftitle={ATLAS document},pdfauthor={The ATLAS Collaboration}}
 
\begin{document}
 
\maketitle
\clearpage

\section{Introduction}
\label{sec:intro}
A new boson~\cite{atlas-higgs, cms-higgs} was discovered in 2012 by the ATLAS and CMS Collaborations.
The observed properties of the particle, such as its couplings to Standard Model (SM) elementary
particles, its spin and its parity, are so far consistent with the predictions
for a SM Higgs boson ($H$)~\cite{HIGG-2013-17, CMS-HIG-14-018, HIGG-2015-07, CMS-HIG-17-031,HIGG-2018-57}.
The mass of this boson was determined by the ATLAS and CMS Collaborations to be
$\mH = 125.09 \pm 0.21\mathrm{(stat)} \pm 0.11\mathrm{(syst)}$~\GeV\ using the LHC Run 1 data set ~\cite{higgs-mass}. Subsequent measurements,  by both collaborations during the LHC Run 2, are published~\cite{HIGG-2016-33,CMS-HIG-16-041,CMS-HIG-19-004} and are consistent with this value.

The SM Higgs boson can decay into \Zg through loop diagrams
and the branching ratio is predicted to be $\brhzg = (1.54\pm0.09)\times 10^{-3}$ at $\mH = 125.09$~\GeV~\cite{deFlorian:2016spz}.
It can differ from the SM value for several scenarios beyond the SM, for example,
if the Higgs boson were a neutral scalar of different origin~\cite{low, low2}, or a composite state~\cite{azatov}.
Different branching ratios are also expected for models with additional colourless charged scalars, leptons
or vector bosons that couple to the Higgs boson, due to their contributions via loop corrections~\cite{carena, extensions, Chen:2013vi}.

Final states where the $Z$ boson decays into electron or muon pairs
can be efficiently triggered and clearly distinguished from background events produced in $pp$ collisions.
In addition, the $Z(\to\ell\ell)\gamma$ ($\ell= e$ or $\mu$) final state can be reconstructed completely
with good invariant mass resolution and relatively small backgrounds.

Previous searches for the $H\to Z(\to\ell\ell)\gamma$ decay  by the ATLAS and CMS
Collaborations use the full $pp$ data sets collected at $\sqrt{s}=7$ and
8~\TeV~\cite{atlas-lowmass,cms-lowmass} and partial data sets collected at 13~\TeV~\cite{ATLASZy-2017,CMSZylow-2018}.
In all cases, no significant excess of events above the expected background is observed around the Higgs boson mass.
Prior to the present study, the ATLAS Collaboration reported an observed
(expected assuming the presence of a SM Higgs boson signal) upper limit on the production cross-section times branching ratio for $pp \to H\to Z\gamma$ of 6.6 (5.2) times the SM prediction at 95\%
confidence level for a Higgs boson with
$\mH = 125.09$~\GeV{} using a sample corresponding to an integrated luminosity of 36.1 \invfb~\cite{ATLASZy-2017}.
The CMS Collaboration reported an observed (expected assuming the presence of a SM Higgs boson signal) upper limit
of 7.4 (6.0) times the SM prediction for a Higgs boson with $\mH = 125$~\GeV{} using a sample corresponding to an integrated luminosity of 35.9 \invfb~\cite{CMSZylow-2018}.

An updated search for decays of the Higgs boson into \Zg with the $Z$
boson decaying into electron or muon pairs is detailed in this letter.
The search uses $pp$ collision data recorded at
\sqtt~with the ATLAS detector at the LHC from 2015 to 2018, corresponding to a total
integrated luminosity of \lumithirteentev~\invfb.
There are important improvements in this analysis compared with the previous one~\cite{ATLASZy-2017},
including an
increase in the size of the data set, an improved event categorisation, and optimised lepton and
photon identification criteria. 
Events with at least one photon and two electrons or muons of opposite charge
are classified into six mutually exclusive categories which are designed
to enhance the sensitivity to the presence of the SM Higgs boson decaying into $Z\gamma$.
The dominant background is the irreducible non-resonant production
of $Z$ bosons together with photons.
A simultaneous fit to the reconstructed $Z\gamma$ invariant mass distributions in all the categories is
performed to extract the overall \htoZg~signal yield.


\section{ATLAS detector and data sample}
\label{sec:samples}
The ATLAS detector~\cite{ATLASDET-2008, ATLAS-TDR-2010-19} at the LHC is a multipurpose particle detector with a forward--backward
symmetric cylindrical geometry and a near 4$\pi$ coverage in solid angle.\footnote{The ATLAS experiment uses a
right-handed coordinate system with its origin at the nominal interaction point (IP) in the centre
of the detector and the $z$-axis along the beam pipe. The $x$-axis points from the IP to the centre of the
LHC ring, and the $y$-axis points upward. Polar coordinates $(r,\phi)$ are used in the transverse plane.
The polar angle ($\theta$) is measured from the positive $z$-axis and the azimuthal angle ($\phi$) is measured from the positive $x$-axis in the transverse plane.
The pseudorapidity is defined in as $\eta=-\ln\tan(\theta/2)$. Angular distance is measured in units of $\Delta R = \sqrt{(\Delta\eta)^2 + (\Delta\phi)^2}$.}
It consists of an inner tracking
detector (ID) surrounded by a thin superconducting solenoid, electromagnetic (EM) and hadronic calorimeters,
and a muon spectrometer (MS) incorporating three large air-core toroidal magnets with eight coils each.

A two-level trigger system~\cite{ATLAS:trigperf} was used during the \sqtt~data-taking period.
The first-level trigger (L1) is implemented
in hardware and uses a subset of the detector information. This is followed by a software-based
high-level trigger which runs algorithms similar to those in the offline reconstruction software, reducing the event rate to
approximately 1~kHz from the maximum L1 rate of 100~kHz.

The events were collected with triggers requiring either one or two electrons or muons in the event.
Due to the increasing luminosity, the transverse momentum ($\pt$) thresholds were increased slightly
during the data-taking periods.
At the highest instantaneous luminosity recorded, the lowest \pt threshold for the single-muon trigger was 26~\GeV.
For the dimuon trigger, asymmetric \pt thresholds of 22~\GeV\ and 8~\GeV\ were used.
The \pt threshold was 26~\GeV\ for the single-electron trigger and 17~\GeV\ for both electrons in the dielectron trigger.
These lowest-threshold triggers were complemented by triggers with higher thresholds but looser lepton identification criteria.
For $H\to Z\gamma$ events that pass the full analysis selection, described in Section~\ref{sec:eventSelection},
the trigger selection is 95.6\% efficient for the $ee\gamma$ final state and 92.2\% efficient
for the $\mu\mu\gamma$ final state. The trigger efficiency has been measured to an accuracy better than 2\%.
After applying trigger and
data quality requirements, the integrated luminosity of the data used in this search corresponds to $\lumithirteentev \pm 2.4$~\invfb.
The average number of $pp$ interactions per bunch crossing (\pileup) ranged from about 13 in 2015 to about 39 in 2018,
with a peak instantaneous luminosity of \peakLumi.
 

\section{Simulation samples}
\label{sec:mcsamples}
Simulated Monte Carlo (MC) events of the signal and dominant backgrounds
are used to optimise the search strategy.
The generated MC events, unless stated otherwise, were processed with the detailed ATLAS detector simulation~\cite{ATLAS-SIM}
based on $\GEANT4$~\cite{Geant4}. The effects of \pileup were modelled by overlaying simulated inelastic $pp$
events over the original hard-scattering event. The pile-up events were generated with \PYTHIAV{8.186}~\cite{Sjostrand:2007gs}
using the NNPDF2.3LO set of parton distribution functions (PDFs)~\cite{Ball:2012cx} and the A3 parameter tune~\cite{ATL-PHYS-PUB-2016-017}.  The MC events were weighted to reproduce
the distribution of the number of interactions per bunch crossing observed in the data.

In order to improve the description of the data, simulated events were corrected to ensure that the efficiencies for the reconstruction and identification of objects match those measured in data. The corrections include those applied to trigger, reconstruction, identification and isolation efficiencies for electrons and muons, identification and selection efficiencies for photons, and selection efficiency for jets~\cite{EGAM-2018-01,Aad:2016jkr}.
Similarly, momentum and energy scale and resolution corrections for simulated objects were also taken into account.

The mass of the Higgs boson for all simulated samples was chosen to be $\mh = 125$~\GeV{} and the
corresponding width is $\Gamma_H = 4.1$~\MeV~\cite{deFlorian:2016spz}.
The samples were normalised with the latest available theoretical calculations of the
corresponding SM production cross-sections at $\mh = 125.09$~\GeV{} via gluon--gluon fusion (\ggF)~\cite{deFlorian:2016spz,Anastasiou:2015ema,Anastasiou:2016cez,Dulat:2018rbf,Harlander:2009mq,Harlander:2009bw,Harlander:2009my,Pak:2009dg,Actis:2008ug,Actis:2008ts,Anastasiou:2008tj,Aglietti:2004nj,Bonetti:2018ukf},
via vector-boson fusion (\VBF)~\cite{deFlorian:2016spz,Ciccolini:2007jr,Ciccolini:2007ec,Bolzoni:2010xr},
in  association with a vector boson (\VH, where $V$ is a $W$ or a $Z$ boson) ~\cite{deFlorian:2016spz,Ciccolini:2003jy,Brein:2003wg,Brein:2011vx,Altenkamp:2012sx,Denner:2014cla,Brein:2012ne,Harlander:2014wda, Harlander:2018yio}
and with a top-quark pair (\ttH) ~\cite{Beenakker:2002nc,Dawson:2003zu, Yu:2014cka,Frixione:2015zaa,deFlorian:2016spz}.
The branching ratio is calculated at leading order in QCD~\cite{Djouadi:1997yw,SpiraQCDHIGGS,Djouadi:2006bz,deFlorian:2016spz,DJOUADI2019214} and it has been shown that higher order QCD corrections have a small impact on the estimated coupling strength~\cite{SPIRA1992350,Bonciani2015,Gehrmann2015}.

The production of the SM Higgs boson
was modelled with the \POWHEGBOXV{v2} Monte Carlo event generator~\cite{Nason:2009ai,Alioli:2010xd,Nason:2004rx,Frixione:2007vw,Hartanto:2015uka},
as described in Ref.~\cite{HIGG-2018-028} and summarised in Table \ref{tab:samples}. \PYTHIAV{8}~\cite{Sjostrand:2007gs} was used to simulate the \htoZg decay as well as to provide parton showering, hadronisation and the underlying event.
Other Higgs boson production processes are not considered as their contributions to the total Higgs boson production cross-section are of the order of $0.1\%$ or less.
All four production modes of the Higgs boson considered contribute to the signal in this analysis and their relative yields were fixed to the SM predictions.
Contributions from $H\rightarrow\mu\mu$ (where the reconstructed photon originates from QED final-state radiation)  were evaluated using samples produced in the same manner and are considered as a potential background in this analysis.
The impact of interference between Higgs bosons decays with the same final-state signature ($H\rightarrow\gamma^*\gamma$, $\gamma^{*}\rightarrow e^{+}e^{-}/\mu^{+}\mu^{-}$ and $H\rightarrow\mu\mu$) is expected to be negligible in the Standard Model~\cite{llyBR2013} and is neglected.

Additional samples of Higgs bosons produced by gluon--gluon fusion are used for studies of theoretical
uncertainties.  A sample with multiple parton interactions disabled is used to study the
uncertainties in the signal acceptance related to the modelling of non-perturbative quantum chromodynamics (QCD) effects.
A sample generated with \MGMCatNLO~\cite{Alwall:2014hca} 
using the NNPDF30 PDF \cite{Butterworth:2015oua, Ball:2014uwa} set, which includes up to two jets at next-to-leading-order (NLO) accuracy in QCD using the FxFx merging scheme ~\cite{Alwall:2014hca,Frederix:2012ps}, is used to study the \ggF acceptance in the analysis categories.

The background in this analysis originates mainly from non-resonant production of a $Z$ boson and a
photon ($Z\gamma$), with a smaller contribution from the production of $Z$
bosons in association with jets ($Z+$jets), with one jet misidentified as
a photon.
 
A large sample of background $Z\gamma$ events was simulated with the
\SHERPAV{v2.2.2}~\cite{Bothmann:2019yzt} generator using a fast simulation of the calorimeter response~\cite{fastsim}. 
It was produced at NLO precision in QCD for up to one additional parton and leading-order (LO) accuracy in QCD for up to three additional partons using  the NNPDF3.0nnlo PDF set~\cite{Ball:2014uwa}. The matrix elements were matched and merged with the \SHERPA parton shower~\cite{Schumann:2007mg,Gleisberg:2008fv}  using the MEPS@LO prescription~\cite{Catani:2001cc,Hoeche:2009rj,Hoeche:2011fd,Hoeche:2012yf}.
 
The electroweak production of $Z\gamma jj$ with jets originating from the fragmentation of partons
arising from electroweak vertices is also considered. It was generated at \LO accuracy in QCD 
using \MGMCatNLOV{2.3.3} with no additional partons in the final state, which is orthogonal to the $Z\gamma$ simulation with Sherpa.
The NNPDF30 \LO PDF set~\cite{Ball:2014uwa} was used for the generation of the events, and the hadronisation, parton shower and the underlying event of the events was modelled using \PYTHIAV{8.212} with the A14 parameter tune~\cite{ATL-PHYS-PUB-2014-021}.

The background originating from $Z$+jets is estimated using a data-driven technique which is validated using a sample simulated with the \POWHEGBOXV{v1} MC generator~\cite{Nason:2004rx,Frixione:2007vw,Alioli:2010xd} at NLO accuracy in QCD.
It was interfaced to \PYTHIAV{8.186} for the modelling of the parton shower, hadronisation and the underlying event with parameters set according to the AZNLO tune~\cite{STDM-2012-23}. The CT10 PDF set~\cite{Lai:2010vv} was used
for the hard-scattering processes, whereas the CTEQ6L1 PDF
set~\cite{Pumplin:2002vw} was used for the parton shower.

\begin{table}[htbp]
\caption{Higgs boson MC samples produced with \POWHEGBOXV{v2} with the techniques
used to generate the event and their precision in $\alpha_\mathrm{s}$ for the event generation (gen.).
The version of \PYTHIAV{8} and the PDF set used for modelling the Higgs bosons decay, parton shower, hadronisation and the underlying event are listed.
The precision of the total cross-section used in the sample normalisation is also reported. The cross-section of the \qqZH sample is normalised to take into account contributions from both \qqZH and \ggZH.  }
\label{tab:samples}
\begin{center} \scriptsize
\begin{tabular}{llllll}
\hline\hline
Process & Technique         & \PYTHIAV{8} version \& tune & PDF set     & QCD (gen.) & Normalisation \\
\hline
\ggF    &  \begin{tabular}{@{}l@{}} \\ NNLOPS~\cite{Hamilton:2013fea,Hamilton:2015nsa}  \\ \& MiNLO~\cite{Bozzi:2005wk,Catani:2007vq,Bagnaschi:2011tu,deFlorian:2011xf,Hamilton:2012np,Campbell:2012am,Hamilton:2012rf}   \end{tabular}  & 8.186, AZNLO \cite{STDM-2012-23} & NNPDF30  \cite{Ball:2014uwa,Butterworth:2015oua} & \begin{tabular}{@{}l@{}}NNLO \end{tabular} & NNNLO (QCD),  NLO (EW)~\cite{Aglietti:2004nj,Actis:2008ug,Actis:2008ts,Harlander:2009mq,Harlander:2009bw,Harlander:2009my,Pak:2009dg,Anastasiou:2008tj,Anastasiou:2015ema,Anastasiou:2016cez,Dulat:2018rbf,Bonetti:2018ukf,Bonetti:2018ukf} \\
\VBF    &  \POWHEG~\cite{Nason:2004rx,Frixione:2007vw,Nason:2009ai,Alioli:2010xd,Hartanto:2015uka}                   & 8.186, AZNLO & NNPDF30 & NLO                      &  NNLO (QCD), NLO (EW) ~\cite{Ciccolini:2007jr,Ciccolini:2007ec,Bolzoni:2010xr} \\
\qqVH   &  \begin{tabular}{@{}l@{}}MiNLO \cite{Cullen:2011ac,Hamilton:2012rf,Luisoni:2013kna}   \end{tabular} & 8.186, AZNLO & NNPDF30 & NLO  & NNLO (QCD), NLO (EW)~\cite{Ciccolini:2003jy,Brein:2003wg,Brein:2011vx,Altenkamp:2012sx,Brein:2012ne,Denner:2014cla,Harlander:2014wda, Harlander:2018yio} \\
\ttH & \POWHEG                 &  8.230, A14  \cite{ATL-PHYS-PUB-2014-021} & NNPDF23~\cite{Ball:2012cx} & NLO  & NNLO (QCD), NLO (EW)  \cite{Beenakker:2002nc,Dawson:2003zu, Yu:2014cka,Frixione:2015zaa}\\
\hline\hline
\end{tabular}
\end{center}
\end{table}
 
 
\section{Event selection, reconstruction and categorisation}
\label{sec:eventSelection}
Events are required to have at least one photon candidate and at least two same-flavour opposite-charge leptons, ($\ell = e, \mu$),
associated with a primary vertex candidate.
This primary vertex candidate is reconstructed from charged particles (tracks) in the ID with transverse
momentum $\pt > 500$~\MeV, and is defined
to be the one with the largest sum of the squared transverse momenta of the associated tracks.
 
Muon candidates are required to have a high-quality track in the ID
or the MS, satisfy the \textit{medium} identification criteria, be within $|\eta|<2.7$ and have $\pt > 10$~\GeV~\cite{Aad:2016jkr}.
Electron and photon candidates are reconstructed from topological clusters of energy deposits in the EM calorimeter,
and in the case of an electron, a track in the ID matched to the cluster~\cite{EGAM-2018-01,PERF-2017-02}.
Electron and photon candidates in the transition region between the barrel and endcap EM calorimeters, $1.37 < |\eta| < 1.52$, are excluded.
Electrons are required to satisfy \textit{loose} likelihood-based identification criteria~\cite{EGAM-2018-01},
have $\pt > 10$~\GeV\ and be within $|\eta|<2.47$, while photon candidates are required to satisfy $\pt > 10$~\GeV, $|\eta|<2.37$ and
the \textit{tight} identification criteria.   Compared with the
previous ATLAS publication~\cite{ATLASZy-2017} the electron identification criteria is looser, improving signal acceptance, and the photon identification has been updated to improve efficiency in the transverse momentum range $10~\GeV-35~\GeV$.
The lepton and photon candidates are also required to be isolated from additional activity in the tracking detector and in the calorimeters.
Contributions to the energy deposited in the calorimeters originating from the underlying events and pile-up are corrected for on an event-by-event basis
using the method described in Refs.~\cite{STDM-2010-08,Cacciari:2009dp,Cacciari_2008}.

In order to ensure that muon and electron candidates originate from the primary vertex, it is required
that the longitudinal impact parameter, $\Delta\zzero$, computed relative to the
primary vertex position, satisfies $\zzsint < 0.5$~mm, where $\theta$ is the polar angle of the track.
Additionally, to suppress leptons from heavy-flavour decays the significance of the
transverse impact parameter $\dzero$ calculated relative to the measured beam-line position must satisfy
$|\dzero| / \sigma_{\dzero} < 3~(5)$ for muons (electrons) where $\sigma_{\dzero}$ is the
uncertainty in $\dzero$ obtained from the track fit.
 
Jets are reconstructed from topological clusters~\cite{Aad:2016upy}
using the anti-$k_t$ algorithm~\cite{Cacciari:2008gp} with a radius parameter of 0.4.
They are required to have $\pt > 25~\GeV$ and $|\eta|<4.4$.
Jets produced in \pileup interactions are suppressed by requiring that those with $\pt <
60~\GeV$ and $|\eta|<2.4$ pass a selection based on a jet vertex tagging algorithm~\cite{ATLAS-JVT},
which is 92\% efficient for jets originating from the hard interaction.

An overlap removal procedure is applied to the selected lepton, photon and jet candidates.
If two electrons share the same track, or the separation between two electron energy clusters
satisfies $|\Delta\eta| < 0.075$ and $|\Delta\phi| < 0.125$, then only the highest-$\pt$ electron
is retained. Electron candidates that fall within
$\Delta R = 0.02$ of a selected muon candidate are also discarded. 
In order to suppress the events arising from QED final-state radiation (FSR),
photon candidates within a $\Delta R = 0.3$ cone around
the leptons of the $Z$ boson candidate are rejected.
Jet--lepton and jet--photon overlap removal are also performed by removing the jet
if its axis is within a cone of size $\Delta R$ = 0.2  around one of the leptons or the photon.

The $Z$ boson candidates are reconstructed from two same-flavour opposite-charge leptons satisfying the selection criteria.
The leptons of the $Z$ boson candidate are additionally required to be consistent with
being accepted by at least one of the triggers that the event passed.
It is required that the \pT of the leptons that are associated with the single-lepton or dilepton triggers is 1~\GeV\ above the trigger threshold.  The invariant mass of the $Z$ boson candidates is required to be between 50~\GeV\ and 101~\GeV\ before applying the mass resolution improvements described in the following.

The resolution of the invariant mass of the  $Z\to\mu\mu$ candidates is
improved by 3\% by correcting the muon momenta for collinear FSR ($\Delta R < 0.15$), using all photons identified in the EM calorimeter~\cite{HIGG-2018-028}.
A constrained kinematic fit is applied to the dilepton invariant mass~\cite{Aad:2014eva} for all $Z$ boson candidates.
This fit uses a line shape modelled by a Breit-Wigner distribution using the world average values for $Z$ bosons mass and width~\cite{Olive:2016xmw} and a single Gaussian to model the lepton momentum response.
It improves the mass resolution by 14\% for the Higgs boson candidates with electrons in the final state
and by 10\% for final states with muons when combined with the FSR correction.
 
After the FSR correction and applying the kinematic fit, $Z$ boson candidates are required to have
an invariant mass within 10~\GeV\ of the $Z$ boson mass, $m_Z = 91.2~\GeV$.
If an event has multiple $Z$ boson candidates which pass all
requirements, the candidate with the mass closest to the $Z$ boson mass is chosen.
Fewer than 1\% of the simulated signal events that pass the final \htoZg
selection have more than two leptons (dominated by events produced via \VH and \ttH) and less than 0.1\% of the signal events have a $Z$ boson
candidate that does not match a true $Z$ boson.
 
The Higgs boson candidate is reconstructed from the $Z$ boson candidate and the highest-$\pt$
photon candidate in the event. The invariant mass of the final-state
particles, $m_{Z\gamma}$ after correction for FSR and by the kinematic fit, is required to satisfy $105~\GeV < \mzg < 160$~\GeV.
Finally, to reduce background contamination and simplify the background modelling,
an additional requirement is placed on the transverse momentum of the photon such that $\pt^{\gamma} / \mzg > 0.12$.
 
For SM $H\to Z(\to\ell\ell)\gamma$ events, the reconstruction and selection efficiency (including
kinematic acceptance) is 20.4\% varying by a maximum of 2\% depending on the production mode.

In order to improve the sensitivity to a $H\to Z\gamma$ signal, the selected events are classified into six mutually exclusive categories
with different expected signal-to-background ratios and mass resolutions.
Categories are defined according to the lepton flavour and event kinematics.
Additionally, a boosted decision tree (BDT) trained to separate VBF signal events from other Higgs boson production modes and
backgrounds is used to define a category of events with at least two jets.
If there are more than two jets in an event, the two highest-\pT jets are used.
The kinematic variables used in the categorisation and as input to the BDT, which have been extended with respect to the
Ref. ~\cite{ATLASZy-2017}, are:
\begin{itemize}
\item The \pT of the highest-\pT jet, $p_\mathrm{T}^{j1}$.
\item The pseudorapidity difference between the two jets, $\Delta\eta_{jj}$.
\item The minimum $\Delta R$ between the $Z$ boson or photon
candidate and either of the two jets, $\Delta R^\mathrm{min}_{\gamma\mathrm{\,or\,}Z,j}$.
\item The invariant mass of the two jets, $m_{jj}$.
\item The absolute value of the difference between the pseudorapidity of the $Z\gamma$
system and the average pseudorapidity of the two jets, $\left( |\eta_{Z\gamma}-(\eta_{j1} + \eta_{j2})/2|\right)$~\cite{Zeppenfeld}.
\item The azimuthal separation between the \Zg system and the system formed by the two jets, $\Delta\Phi_{Z\gamma,jj}$.
\item  The azimuthal angle between the dilepton system and the photon, $\Delta\Phi_{Z,\gamma}$.
\item The component, $\ptt$, of the transverse momentum of the \Zg system,
that is perpendicular to the difference of the 3-momenta of the $Z$ boson and the photon
candidate ($\ptt = |\vec{p}_\mathrm{T}^{Z\gamma} \times \hat{t} |$, where $\hat{t} = (\vec{p}_\mathrm{T}^{Z}-\vec{p}_\mathrm{T}^{\gamma})/|\vec{p}_\mathrm{T}^{Z}-\vec{p}_\mathrm{T}^{\gamma}|$ ).
This quantity is strongly correlated with the transverse momentum of the $Z\gamma$ system,
but has better experimental resolution~\cite{Ackerstaff:1997rc,VESTERINEN2009432}.
\end{itemize}

Events with two or more jets with $\Delta\eta_{jj} > 2$ that have a BDT output score larger than 0.87
are classified into a \textit{VBF-enriched} category.
Of the remaining events, those that satisfy the requirement on
$\pt^\gamma / m_{Z\gamma} > 0.4$ are classified into a \textit{High relative $\pt$} category
while the others are separated into four categories depending on the lepton
flavour and a cut at $\ptt=40~\GeV$.  The boundaries of the categories are
selected to maximise the expected signal significance.

VBF events are estimated to constitute 72\% of the signal in the \textit{VBF-enriched category}.
The \textit{High relative \pT} and \textit{High \ptt} categories are expected to be enriched in
\VBF, \VH and \ttH events as these production modes have on average higher Higgs boson \pT
than \ggF production.
Because the continuum \Zg background has on average lower Higgs boson candidate \pT than the signal,
the signal-to-background ratio is expected to be higher in these categories
than in the other categories, as shown in Table~\ref{tab:SBtable}.
The table also summarises the observed number of events in data in the \Zg mass range of 105-160 \GeV\
and the expected number of signal ($S_{68}$) and background ($B_{68}$) events in a \mzg window containing 68\% of the expected signal.
In addition the width of the window containing 68\% of the SM signal ($w_{68}$), which quantifies the mass resolution
and is defined as the difference between the 84th and the 16th percentile of the signal mass distribution, is reported.
$B_{68}$ is estimated from fits to the data using the background models described in Section~\ref{sec:signalBkgModel}.
The categorisation improves the expected sensitivity, which is defined as $S_{68}/\sqrt{S_{68}+B_{68}}$, by approximately 50\%.

\begin{table}[htbp]
\caption{The number of data events selected in each category and in the \Zg mass range of 105--160~\GeV. In addition, the following numbers are given:
the expected number of Higgs boson signal events in an interval around the peak position for a signal
of $\mH=125.09~\GeV$
containing 68\% of the SM signal ($S_{68}$), the mass resolution quantified by the width of the $S_{68}$ interval
($w_{68}$) defined by the difference between the 84th and the 16th percentile of the signal mass distribution, the background in the $S_{68}$ interval ($B_{68}$) is estimated from fits to the data using the background models described in Section~\ref{sec:signalBkgModel}, the observed number of events in the $S_{68}$ interval ($N_{68}$),
the expected signal-to-background ratio in the $S_{68}$ window ($S_{68}/B_{68}$),
and the expected significance estimate defined as $S_{68}/\sqrt{S_{68}+B_{68}}$. The final row of the table displays the expected number of events for an analysis performed in a single inclusive category calculated by summing the number of events in each individual category.}
\label{tab:SBtable}
\begin{center} \scriptsize
\begin{tabular}{lrrrrrcc}
\hline\hline
Category  & Events &  $S_{68}$  &  $B_{68}$~~ &  $N_{68}$~~ & $w_{68}$ [\GeV] & $S_{68}/B_{68}$ $[10^{-2}$] & $S_{68}/\sqrt{S_{68}+B_{68}}$\\
\hline
VBF-enriched                     & 194    & 2.7  & 16.7 &  17 & 3.7 & 16.2~~ & 0.60 \\
High relative $\pt$              & 2276   & 7.6  & 108.5  & 118 & 3.7 & 7.0  & 0.70  \\
High $p_{\mathrm{T}t}$ $ee$      & 5567   & 9.9  & 474.7  & 498 & 3.8 & 2.1  & 0.45 \\
Low $p_{\mathrm{T}t}$ $ee$       & 76\,679  & 34.5& 6418.6& 6505 & 4.1 & 0.5  & 0.43 \\
High $p_{\mathrm{T}t}$ $\mu\mu$  & 6979   & 12.0 & 634.4  & 632 & 3.9 & 1.9  & 0.47 \\
Low $p_{\mathrm{T}t}$ $\mu\mu$   & 100\,876 & 43.5 & 8506.9 & 8491 & 4.0 & 0.5  & 0.47 \\
\hline
Inclusive  & 192\,571 & 110.2 &  16159.8 & 16261& 4.0 & 0.7  & 0.86 \\
\hline\hline
\end{tabular}
\end{center}
\end{table}

 
\section{Signal and background modelling}
\label{sec:signalBkgModel}
 
The signal and background yields are extracted from a fit to the
$m_{Z\gamma}$ distribution observed in data by assuming parametric models for both the signal and backgrounds.
For the signal, the expected acceptance and parameters that describe the shape are obtained from simulated signal samples in the same manner as Ref.~\cite{ATLASZy-2017}.
For the background, the models are chosen using simulated background
samples and the values of their parameters are determined by a fit to the mass spectra measured in data.

The signal mass distribution for the Higgs boson decay into $Z\gamma$ is well modelled
by a double-sided Crystal Ball (DSCB)
function (a Gaussian function with power-law tails on both sides)~\cite{dscb1, dscb2}.
The peak position and width of the Gaussian component are represented by $\mu_\mathrm{CB}$ and $\sigma_\mathrm{CB}$, respectively.
The parameters of the DSCB are determined in each
category by performing a maximum-likelihood fit to the signal MC samples.
 
The parametric model used to describe the background \mzg distribution is selected using
a template that is constructed from the simulated \zgamma and electroweak $Z\gamma jj$ events, and a \zjets~contribution derived from data.  The simulated \zgamma events use a fast simulation of the calorimeter response which has been confirmed to produce a \mzg distribution compatible, in each category, with \zgamma events simulated with the detailed simulation.
The shape of the \mzg distribution for \zjets~events is constructed for each category from a data control region
defined by requiring the photon candidate to not satisfy the tight identification criteria but still pass a looser identification criteria.
To smooth the statistical fluctuations of the \mzg distribution for \zjets~events,
an analytic function is fitted to the ratio of the \mzg distributions for \zjets~and \zgamma~events, in the \mzg range of 105--160~\GeV.
The smoothed \mzg distribution for \zjets~events is constructed by multiplying the \zgamma~distribution by the fitted ratio.
The \zjets~and \zgamma~distributions have similar shapes, allowing a parametric function to be used to fit the ratio\footnote{ The functional form used is $(1-x^{1/3})^{f_0}x^{ p_0 + p_1\log(x)}$ where $x=m_{Z\gamma}/\sqrt{\mathrm{s}}$\,($\sqrt{\mathrm{s}} = 13\, \TeV$) where $f_i$ and $p_i$ are the free parameters of the model.}.
The uncertainty in the fitted ratio is found to have negligible impact on the background template.
 
The background composition in each category is determined using data and confirms the dominance of the \Zg process over other backgrounds where jets are misidentified as photons.
A two-dimensional sideband technique~\cite{Aad:2010sp,ATLASZy-2017} based on the track and calorimeter isolation of the photon
candidate and whether the photon candidate satisfies the tight or the looser identification criteria is performed in each category.
The fraction of \zgamma events with genuine photons (including the $Z\gamma jj$ contribution), inclusive over categories, is $0.78^{+0.04}_{-0.09}$.
 
To construct the final background template for each category, the normalisation of the electroweak $Z\gamma jj$ events is based on the predicted cross-section while the combined \zgamma and \zjets~distribution is normalised to the number
of data events in that category after subtracting the expected number of electroweak $Z\gamma jj$ events.
The background template is treated as a representative Asimov dataset~\cite{stat} when choosing the functional form used to model it.
 
The functional form used to model the background is selected from the following families of functions:
Bernstein polynomials,
exponential polynomial functions ($\mathrm{e}^{(\Sigma^{N}_{i=0} p_{i}m_{Z\gamma}^i)}$),
a sum of power functions ($\Sigma^{N}_{i=1} f_i m_{Z\gamma}^{p_i}$), and a class of functions given by $(1-x^{1/3})^fx^{\Sigma^{N}_{i=0} p_i\log(x)^i}$ where $x=m_{Z\gamma}/\sqrt{\mathrm{s}}$ ($\sqrt{\mathrm{s}} = 13\, \TeV$) where $f_i$ and $p_i$ are the free parameters of the models. The choice of analytical model of the background and the \mzg~range used for the final fit
is optimised in each category using the background template mentioned earlier.  The optimisation
procedure, which has been updated  when compared to Ref.~\cite{ATLASZy-2017}, includes a limit on the amount of bias
in the extracted signal yield (also referred to as spurious signal) and a requirement on the fit
quality, and it prefers models with fewer free parameters to maximise the statistical sensitivity
to the expected signal.
For each category used in the analysis, the bias due to the spurious signal
is estimated by performing a signal-plus-background fit to the
\mzg background-only distribution estimated as explained above, with $m_H$ varied between 123~\GeV\ and 127~\GeV.
The maximum number of signal events derived from these fits in each category constitutes the spurious-signal systematic uncertainty.
A requirement that the spurious signal be less than 50\%
of the expected statistical uncertainty in the signal yield is applied when selecting the background modelling function.
Stricter requirements on the spurious signal are not possible due to the statistical uncertainty in the \mzg background template.
In addition, the $\chi^{2}$ probability of the background-only fit is required to be larger than $1\%$.
The fit range is optimised in each analysis category by varying the lower and upper bounds
in 5~\GeV\ steps within the ranges 105--115~\GeV\ and 140--160~\GeV, respectively.
The optimal fit range and function are selected to achieve the highest signal significance while fitting the expected mass distribution of background plus SM signal.
The significance evaluation also includes the spurious-signal systematic uncertainty.
The selected background functional form and fit range in each category are detailed in Table~\ref{tab:functions}.
 
\begin{table}[!h]
\caption{ The selected background function and fit range in each analysis category.}
\label{tab:functions}
\centering  
\begin{tabular}{llc}
\hline
\hline
Category    & Function Type & Fit range [\GeV] \\ [0.5ex]
\hline
VBF-enriched & Second-order power function& 110--155 \\
High relative $\pt$ & Second-order exponential polynomial& 105--155 \\
$ee$ high $\ptt$ & Second-order Bernstein polynomial& 115--145 \\
$ee$ low $\ptt$ & Second-order exponential polynomial& 115--160 \\
$\mu\mu$ high $\ptt$ & Third-order Bernstein polynomial& 115--160 \\
$\mu\mu$ low $\ptt$& Third-order Bernstein polynomial& 115--160 \\
\hline
\hline
\end{tabular}
 
\end{table}

 
\section{Systematic uncertainties}
\label{sec:sys}
The dominant experimental uncertainty in the signal yield is the spurious signal from the choice of background model.
The spurious signal corresponds to as much as 50\% of the statistical error in the expected signal yield per category, due to a limited number of simulated background events.
It introduces a 28\% systematic uncertainty in the signal strength, defined as the ratio of the observed to expected signal yield; however, it only increases the
total uncertainty in the expected signal strength by 5.6\% because of the large statistical uncertainty.
The contribution from $H\rightarrow\mu\mu$ is estimated using simulated events and amounts to about 1.7\% inclusively, and up to 3.3\% in individual categories. An uncertainty in this contribution taken from the latest ATLAS measurement ~\cite{HIGG-2016-10} results in an uncertainty of
2.1\% in the expected signal yield.
The combined uncertainty in the signal yield in any category due to the reconstruction, identification, isolation, and trigger efficiency
measurements~\cite{EGAM-2018-01,Aad:2016jkr} is no more than 2.6\%, 2.4\% and 1.6\% for photons, electrons and muons, respectively. \Pileup also affects the lepton and photon identification efficiency but contributes a negligible amount to the uncertainty (0.2\%).
The uncertainty in the combined 2015--2018 integrated luminosity is 1.7\%~\cite{ATLAS-CONF-2019-021,LUCID2}.

The theoretical uncertainties in the predicted signal yield originate from uncertainties in the predicted branching ratio (5.7\%)~\cite{deFlorian:2016spz} as well as from uncertainties in the modelling of the production cross-section and kinematics of the Higgs boson due to missing higher-order QCD calculations (5.3\%), that are dominated by uncertainties in the QCD renormalisation and factorisation scales (5.2\%).
Smaller effects originate from the parton shower modelling uncertainty (1.3\%), PDFs (2.5\%) and $\alpha_\mathrm{s}$ (1.9\%).
The uncertainties in the Higgs boson event kinematics due to missing higher-order QCD calculations impact the distribution of signal events amongst the analysis categories.  They are evaluated using an extension of the Stewart--Tackmann method~\cite{Stewart:2011cf, deFlorian:2016spz}, based on inputs from Refs.~\cite{Campbell:2010ff,  Gangal:2013nxa, Grazzini:2013mca}. Details of how the uncertainty in the acceptance of \ggF events in the VBF-enriched category and all other categories is evaluated can be found in Refs.~\cite{HIGG-2016-21} and~\cite{ATLASZy-2017}, respectively.
Additionally, to account for the uncertainties in the modelling of jet kinematics in \ggF events the category acceptance is compared with the acceptance derived from the \MGMCatNLO sample.
The Higgs boson and jet kinematics particularly affect the \ggF signal acceptance in the \VBF-enriched category (37\%) and High relative \pT category (21\%).
The effect of parton shower modelling, PDFs and $\alpha_\mathrm{s}$ on the distribution of signal events amongst the analysis categories is less than 11\%, 1\% and 2\%, respectively.
The expected signal yield in the VBF category is also affected (14\%) by the jet energy scale, jet energy resolution and vertex tagging efficiency~\cite{PERF-2016-04}.
 
The uncertainty in the modelling of the signal shape varies between analysis channels. The uncertainty in the mass resolution ($\sigma_\mathrm{CB}$) is dominated by the uncertainty in the electron and photon energy resolution ($<3.4\%$) and in the muon momentum resolution ($<3.6\%$).  The uncertainty in the signal position ($\mu_\mathrm{CB}$) from the uncertainty in electron, photon and muon calibration ($<0.15\%$) is less than the uncertainty in the assumed Higgs boson mass of $0.19\%$~\cite{higgs-mass}.  The impact of the signal model uncertainty on the signal strength is less than $2\%$.
 

\section{Results}
\label{sec:result}
 
A profile-likelihood-ratio test statistic~\cite{stat} is used to search for
a localised excess of events above the expected background by performing a fit to the \mzg spectra in the various categories.
In the same manner as was done
in previous searches for \htoZg~\cite{ATLASZy-2017},
the likelihood is built from the product of Poisson probability terms across all
categories with two contributions: non-resonant background, and Higgs boson signal.
The likelihood includes terms for the systematic uncertainties discussed in Section \ref{sec:sys} implemented as nuisance parameters. The
nuisance parameters describe the systematic uncertainties, which are parameterised as Gaussian or log-normal priors and are correlated across analysis categories where appropriate.
Upper limits are set on the Higgs boson production cross-section at 95\% confidence level (CL)
using the modified frequentist formalism~\cite{cls}. The results are derived using closed-form asymptotic formulae~\cite{stat}.
 
\begin{figure}[htb]
\begin{center}
\begin{tabular}{cc}
\includegraphics[width=0.365\textwidth]{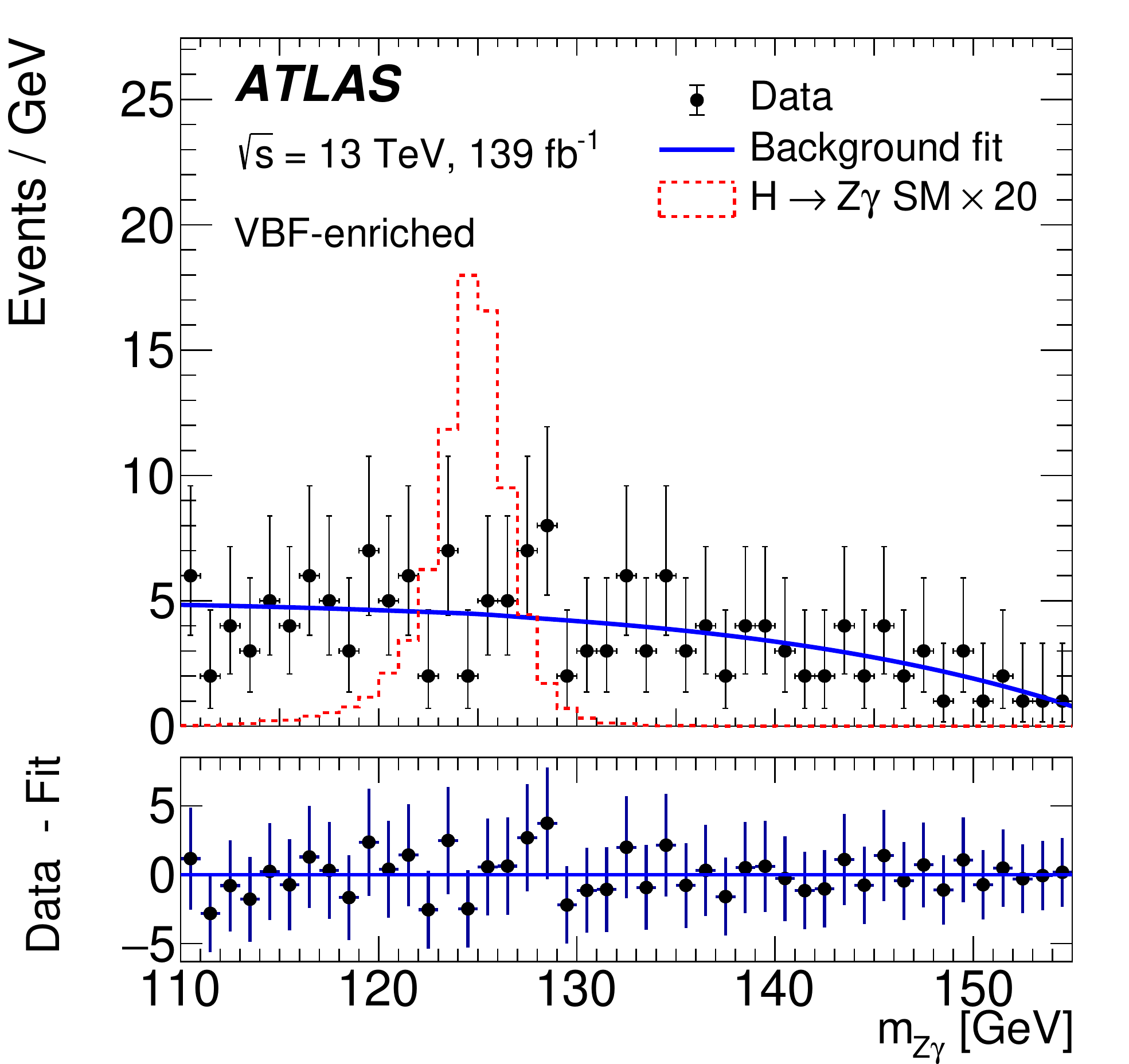} &
\includegraphics[width=0.365\textwidth]{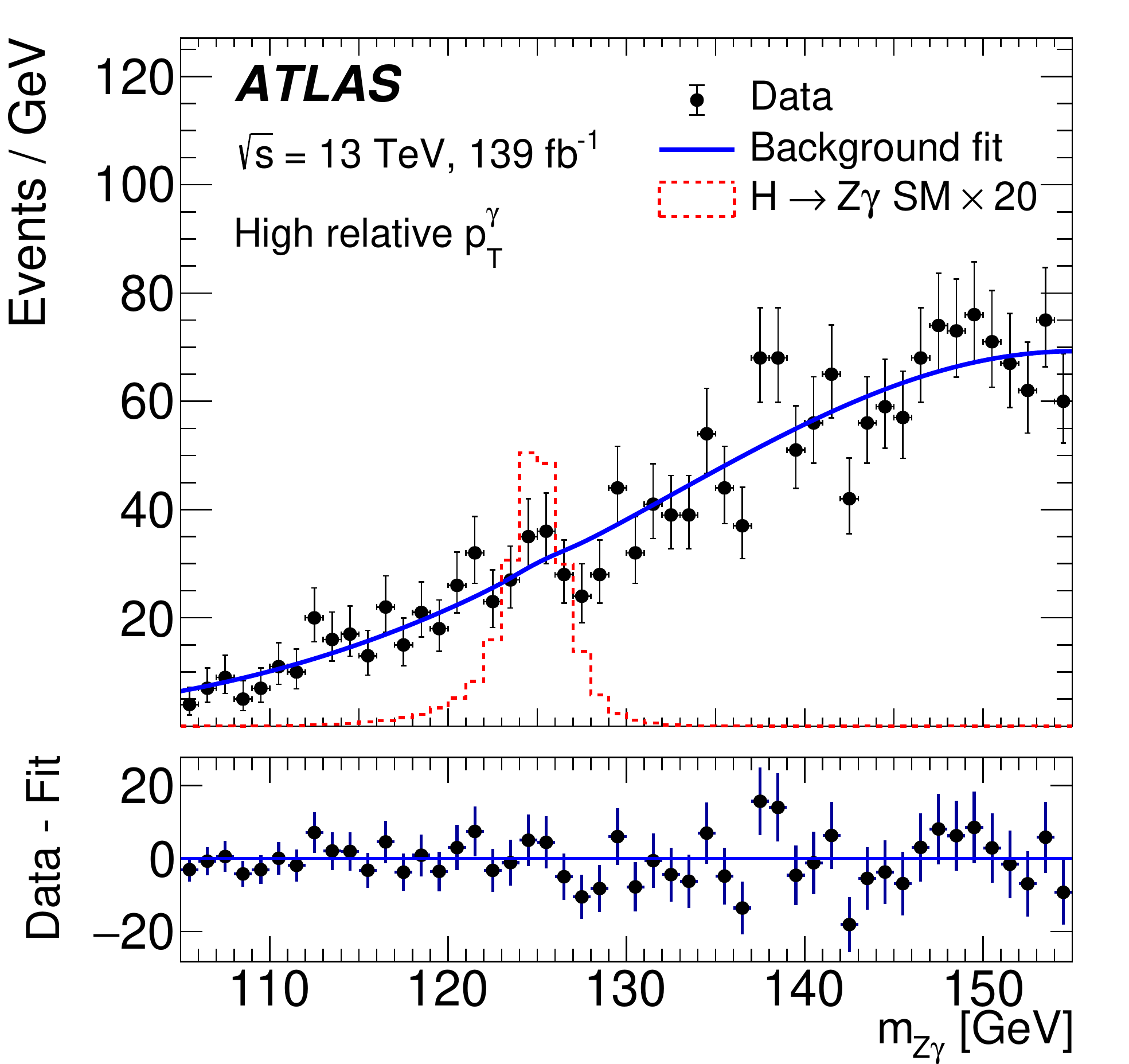} \\
(a) & (b) \\
\includegraphics[width=0.365\textwidth]{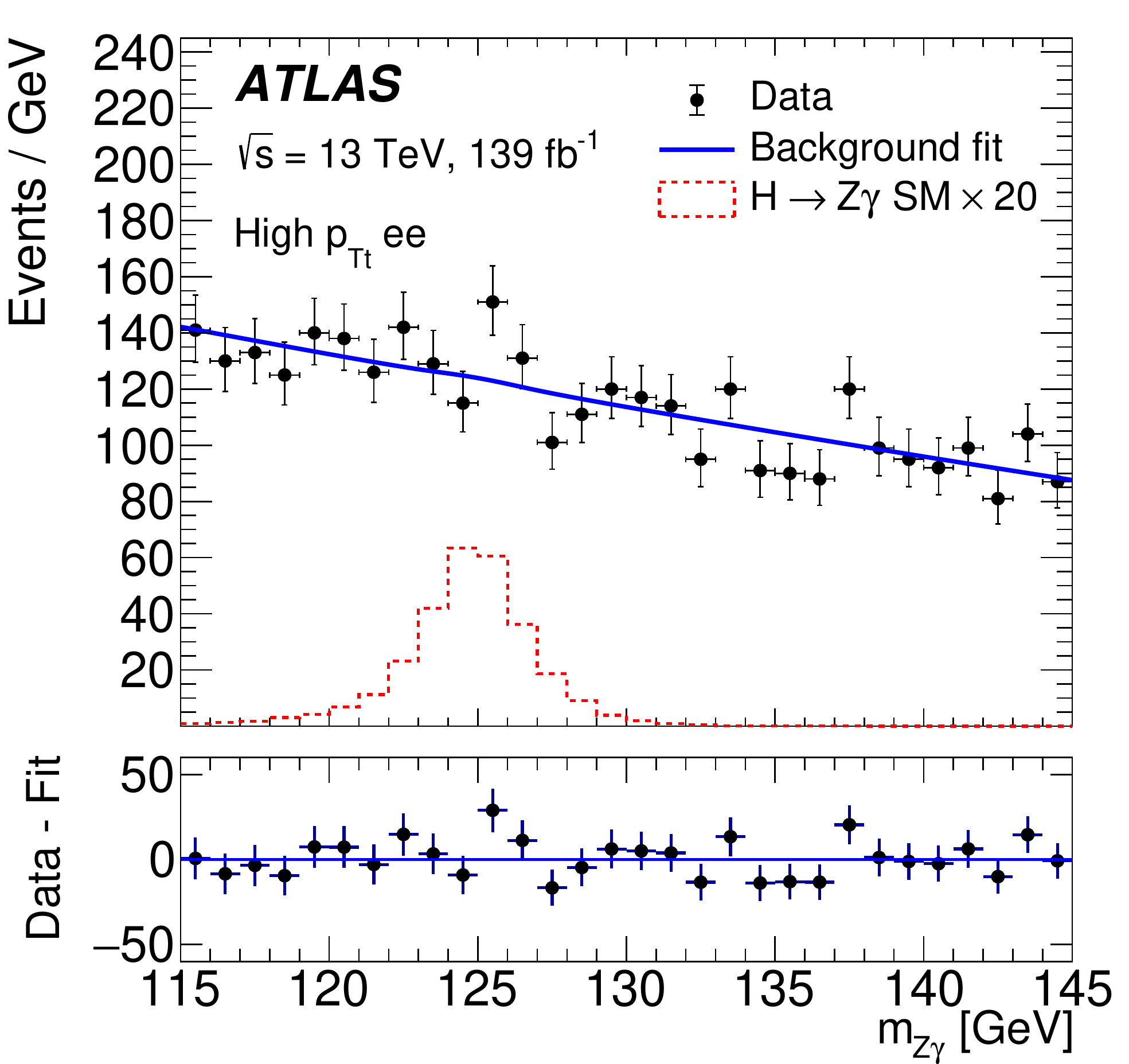} &
\includegraphics[width=0.365\textwidth]{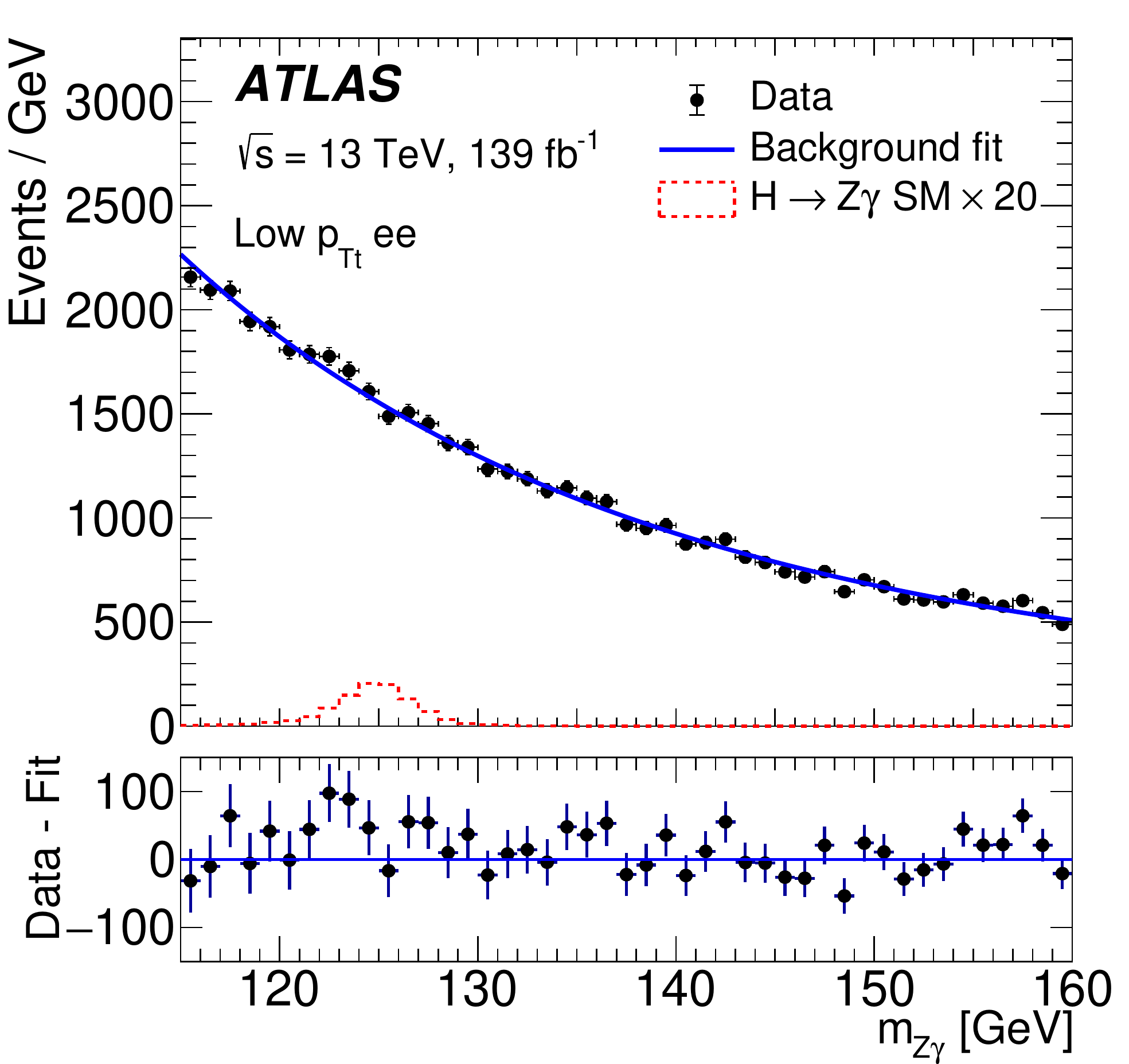} \\
(c) & (d) \\
\includegraphics[width=0.365\textwidth]{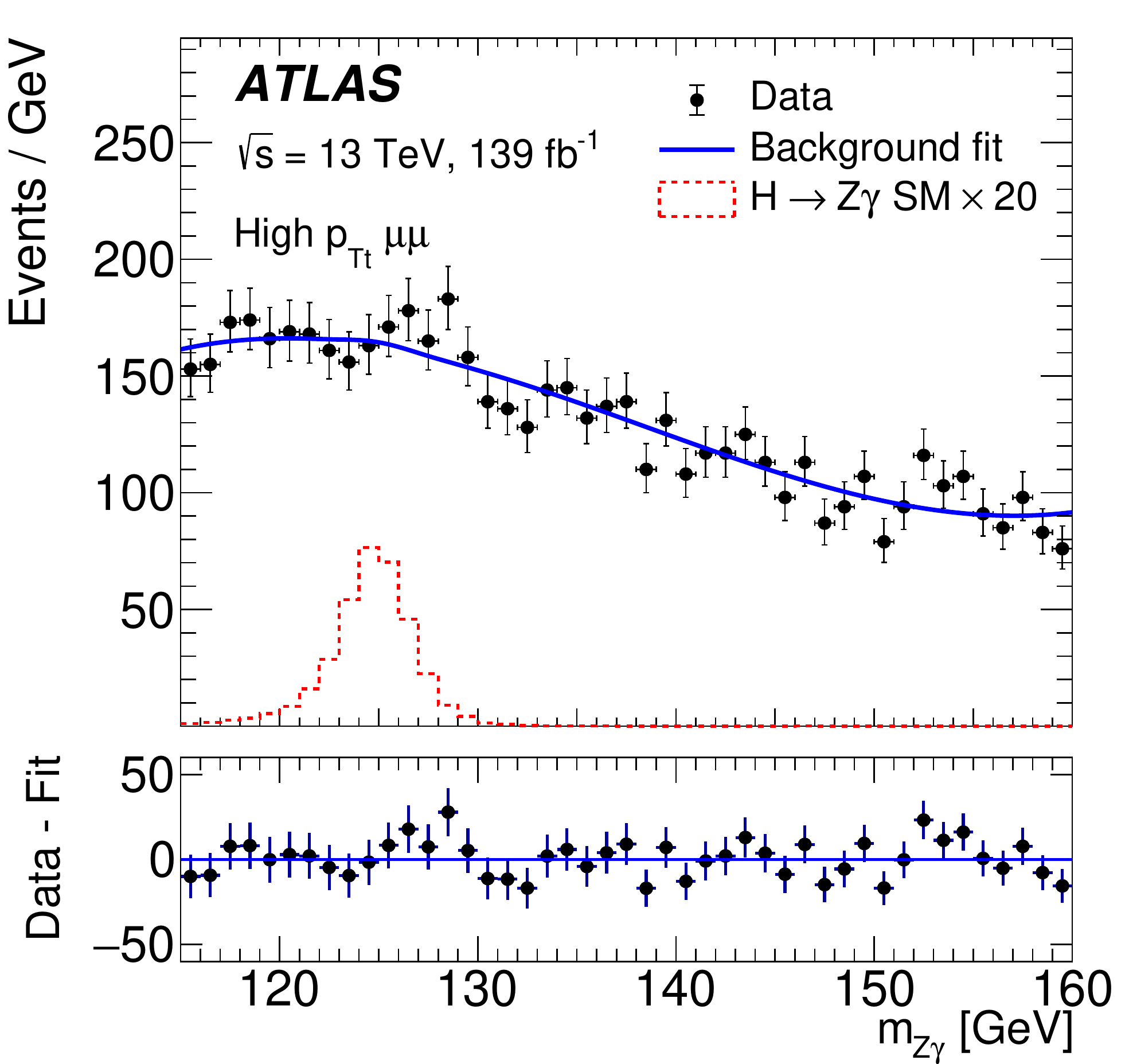} &
\includegraphics[width=0.365\textwidth]{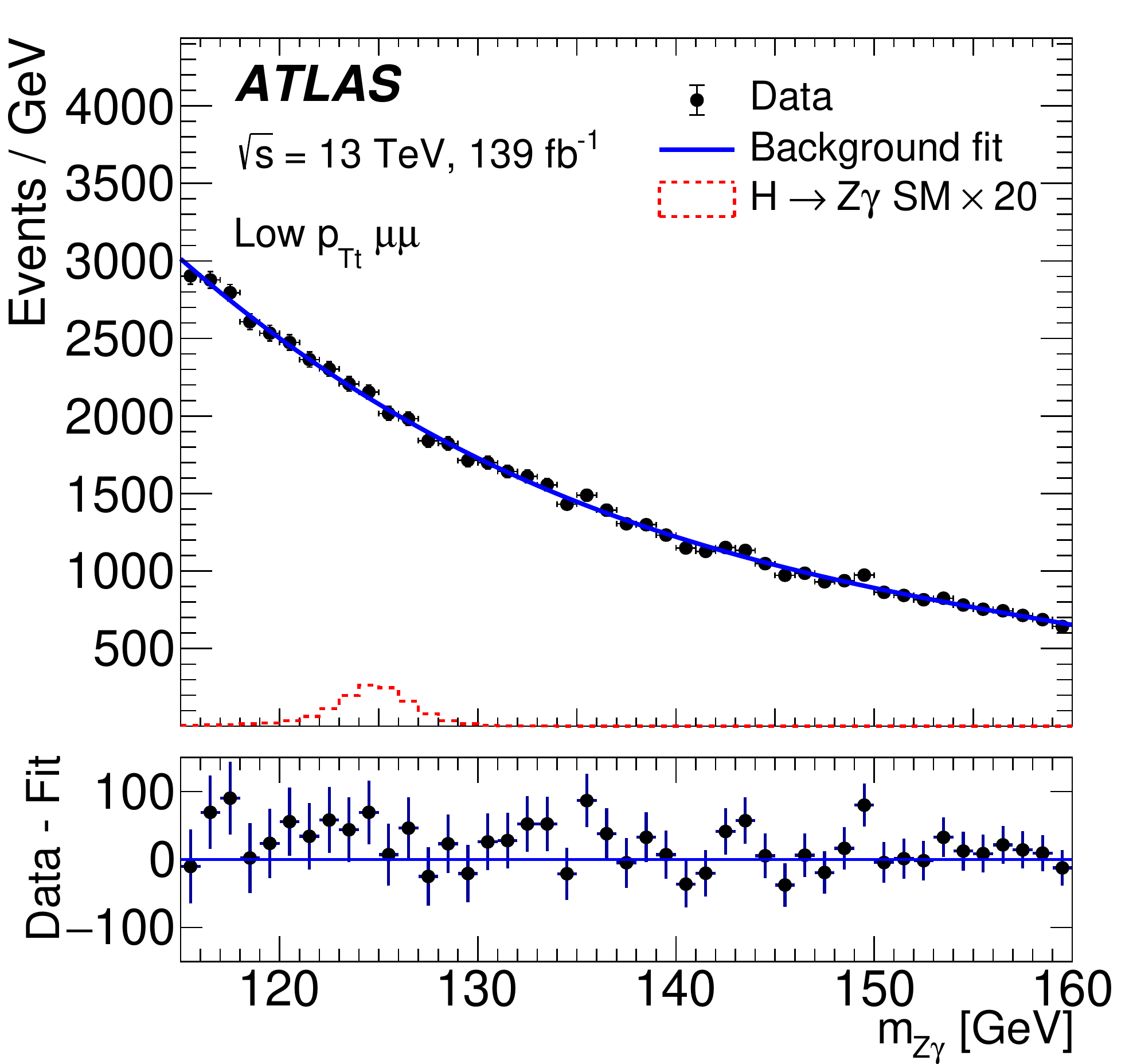} \\
(e) & (f) \\
\end{tabular}
\end{center}
\caption{The $Z\gamma$ invariant mass ($m_{Z\gamma}$) distributions of events satisfying the
$H\to Z\gamma$ selection in data for the six event categories:
(a) VBF-enriched, (b) High relative \ptg, (c) High \ptt $ee$,
(d) Low \ptt $ee$, (e) High \ptt $\mu\mu$ , and (f)  Low \ptt $\mu\mu$.
The black points represent data. The error bars represent only the statistical uncertainty of the data.
The solid blue lines show the
background-only fits to the data, performed independently in each category.
The dashed red histogram corresponds to the expected signal
for a SM Higgs boson with $\mH = 125$~\GeV\
decaying into $Z\gamma$ multiplied by a factor of 20.
The bottom part of the figures shows the residuals of the data with respect to the background-only fit.
}
\label{fig:mllg-low-mass}
\end{figure}

\begin{figure}[htb]
\begin{center}
\includegraphics[width=0.65\textwidth]{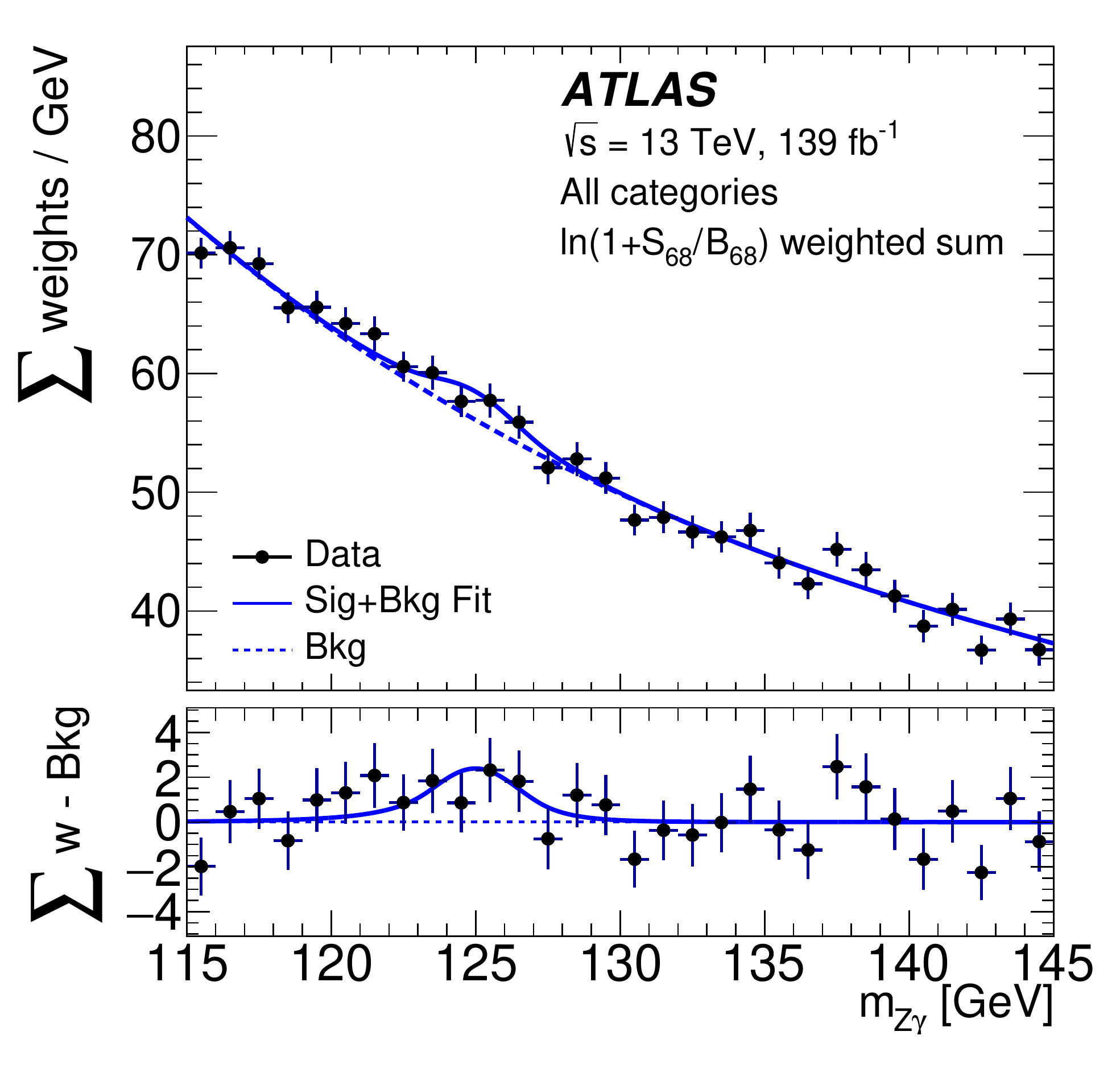}
\end{center}
\caption{Weighted $Z\gamma$ invariant mass ($m_{Z\gamma}$) distribution of events satisfying the
$H\to Z\gamma$ selection in data. The black points represent data.  The error bars represent only the statistical uncertainty of the data.
Events are weighted by $\ln(1+S_{68}/B_{68})$, where $S_{68}$ and $B_{68}$ are the expected signal and background events in a $m_{Z\gamma}$ window containing 68\% of the expected signal.
The solid blue curve shows the combined fitted signal-plus-background model when fitting all analysis categories simultaneously, the dashed line shows the model of the background component.
}
\label{fig:mllg-soverb}
\end{figure}

The invariant mass distributions of the \Zg
events for the various categories are shown in
Figure~\ref{fig:mllg-low-mass} with the background-only fit
superimposed.
The expected Higgs boson signal normalised to 20 times the SM prediction for \mH=125~\GeV\ is also shown.
At \mH=125.09~\GeV, the observed (expected with a SM Higgs boson) $p$-value under the background hypothesis is \lowmassmHpz ~(\lowmassmHpzExpSMHiggs), which corresponds to a significance of \lowmassmHsignificance ~(\lowmassmHsignificanceExpSMHiggs).
A weighted sum of all categories with the fitted signal-plus-background model superimposed is shown in Figure~\ref{fig:mllg-soverb}.
The events are weighted by $\ln(1+S_{68}/B_{68})$, where $S_{68}$ and $B_{68}$ are defined in Section~\ref{sec:eventSelection}.
 
The best-fit value for the \htoZg signal strength, defined as the ratio of the observed to the predicted SM signal yield,
is found to be $2.0\pm0.9(\mathrm{stat.})^{+0.4}_{-0.3}(\mathrm{syst.})=2.0^{+1.0}_{-0.9}(\mathrm{tot.})$ with an expected value of $1.0\pm0.8(\mathrm{stat.})\pm0.3(\mathrm{syst.})$ assuming the presence of the SM Higgs boson.
The measured signal strength amongst all categories are compatible within their total uncertainties.  The largest measured signal strength is $4.7^{+3.0}_{-2.7}(\mathrm{tot.})$ in the High $\ptt$ $ee$ category.
The total uncertainty is dominated by the statistical component from the data. The systematic component of the total uncertainty is dominated by the spurious-signal uncertainties.
 
The observed 95\% CL limit on the \htoZg signal strength is found to be \mhObserved~times the SM prediction compared  with an expected value of
\mhExpected~(\mhExpectedSMHiggs) assuming no (SM) Higgs boson decays into $Z\gamma$.
The observed upper limit on \sighbr~is \mhObservedxsBr~at 95\% CL.
Assuming the SM Higgs boson production cross-section, the upper limit at 95\% CL on
$B (H \rightarrow Z \gamma)$ is found to be 0.55\%.
 
This result represents an improvement of about a factor of 2.4 in expected sensitivity compared with the
previous ATLAS publication ~\cite{ATLASZy-2017}. Of this improvement, a factor of approximately two is due to the larger analysed
dataset and the additional 20\% improvement can be attributed to the improvements in the analysis.

\FloatBarrier

\section{Conclusion}
\label{sec:conclusion}
A search for \Zg decays of the SM Higgs boson
in \lumithirteentev~fb$^{-1}$~of $pp$ collisions at \sqtt~is performed with the ATLAS experiment at the LHC.
The observed data are consistent with the expected background with a $p$-value of 1.3\%, while the
expected p-value in the presence of a SM Higgs boson is 12.3\%. These $p$-values correspond to a
significance of 2.2 and 1.2 standard deviations, respectively.
The observed 95\% CL upper
limit on the \sighbr~is \mhObserved~times the SM prediction for a Higgs boson mass of 125.09 GeV.
The expected limit on \sighbr~ assuming either no Higgs boson decay into \Zg or the presence of the SM Higgs boson decay is \mhExpected~and \mhExpectedSMHiggs~times the SM prediction, respectively.
The best-fit value for the signal yield normalised to the SM prediction is $2.0^{+1.0}_{-0.9}$
where the statistical component of the uncertainty is dominant.
 

\section*{Acknowledgements}
 

We thank CERN for the very successful operation of the LHC, as well as the
support staff from our institutions without whom ATLAS could not be
operated efficiently.
 
We acknowledge the support of ANPCyT, Argentina; YerPhI, Armenia; ARC, Australia; BMWFW and FWF, Austria; ANAS, Azerbaijan; SSTC, Belarus; CNPq and FAPESP, Brazil; NSERC, NRC and CFI, Canada; CERN; CONICYT, Chile; CAS, MOST and NSFC, China; COLCIENCIAS, Colombia; MSMT CR, MPO CR and VSC CR, Czech Republic; DNRF and DNSRC, Denmark; IN2P3-CNRS and CEA-DRF/IRFU, France; SRNSFG, Georgia; BMBF, HGF and MPG, Germany; GSRT, Greece; RGC and Hong Kong SAR, China; ISF and Benoziyo Center, Israel; INFN, Italy; MEXT and JSPS, Japan; CNRST, Morocco; NWO, Netherlands; RCN, Norway; MNiSW and NCN, Poland; FCT, Portugal; MNE/IFA, Romania; MES of Russia and NRC KI, Russia Federation; JINR; MESTD, Serbia; MSSR, Slovakia; ARRS and MIZ\v{S}, Slovenia; DST/NRF, South Africa; MINECO, Spain; SRC and Wallenberg Foundation, Sweden; SERI, SNSF and Cantons of Bern and Geneva, Switzerland; MOST, Taiwan; TAEK, Turkey; STFC, United Kingdom; DOE and NSF, United States of America. In addition, individual groups and members have received support from BCKDF, CANARIE, Compute Canada and CRC, Canada; ERC, ERDF, Horizon 2020, Marie Sk{\l}odowska-Curie Actions and COST, European Union; Investissements d'Avenir Labex, Investissements d'Avenir Idex and ANR, France; DFG and AvH Foundation, Germany; Herakleitos, Thales and Aristeia programmes co-financed by EU-ESF and the Greek NSRF, Greece; BSF-NSF and GIF, Israel; CERCA Programme Generalitat de Catalunya and PROMETEO Programme Generalitat Valenciana, Spain; G\"{o}ran Gustafssons Stiftelse, Sweden; The Royal Society and Leverhulme Trust, United Kingdom.
 
The crucial computing support from all WLCG partners is acknowledged gratefully, in particular from CERN, the ATLAS Tier-1 facilities at TRIUMF (Canada), NDGF (Denmark, Norway, Sweden), CC-IN2P3 (France), KIT/GridKA (Germany), INFN-CNAF (Italy), NL-T1 (Netherlands), PIC (Spain), ASGC (Taiwan), RAL (UK) and BNL (USA), the Tier-2 facilities worldwide and large non-WLCG resource providers. Major contributors of computing resources are listed in Ref.~\cite{ATL-GEN-PUB-2016-002}.
 

\printbibliography
 
\clearpage

\clearpage 
 
\begin{flushleft}
\hypersetup{urlcolor=black}
{\Large The ATLAS Collaboration}

\bigskip

\AtlasOrcid[0000-0002-6665-4934]{G.~Aad}$^\textrm{\scriptsize 102}$,    
\AtlasOrcid[0000-0002-5888-2734]{B.~Abbott}$^\textrm{\scriptsize 128}$,    
\AtlasOrcid{D.C.~Abbott}$^\textrm{\scriptsize 103}$,    
\AtlasOrcid[0000-0002-2788-3822]{A.~Abed~Abud}$^\textrm{\scriptsize 36}$,    
\AtlasOrcid[0000-0002-1002-1652]{K.~Abeling}$^\textrm{\scriptsize 53}$,    
\AtlasOrcid[0000-0002-2987-4006]{D.K.~Abhayasinghe}$^\textrm{\scriptsize 94}$,    
\AtlasOrcid[0000-0002-8496-9294]{S.H.~Abidi}$^\textrm{\scriptsize 166}$,    
\AtlasOrcid[0000-0002-8279-9324]{O.S.~AbouZeid}$^\textrm{\scriptsize 40}$,    
\AtlasOrcid{N.L.~Abraham}$^\textrm{\scriptsize 155}$,    
\AtlasOrcid[0000-0001-5329-6640]{H.~Abramowicz}$^\textrm{\scriptsize 160}$
% ...
\bigskip
\\

% ...
$^{36}$CERN, Geneva; Switzerland.\\
% ...
$^{40}$Niels Bohr Institute, University of Copenhagen, Copenhagen; Denmark.\\
% ...
$^{53}$II. Physikalisches Institut, Georg-August-Universit\"{a}t G\"ottingen, G\"ottingen; Germany.\\
% ...
$^{94}$Department of Physics, Royal Holloway University of London, Egham; United Kingdom.\\
% ...
$^{102}$CPPM, Aix-Marseille Universit\'e, CNRS/IN2P3, Marseille; France.\\
$^{103}$Department of Physics, University of Massachusetts, Amherst MA; United States of America.\\
% ...
$^{128}$Homer L. Dodge Department of Physics and Astronomy, University of Oklahoma, Norman OK; United States of America.\\
% ...
$^{155}$Department of Physics and Astronomy, University of Sussex, Brighton; United Kingdom.\\
% ...
$^{160}$Raymond and Beverly Sackler School of Physics and Astronomy, Tel Aviv University, Tel Aviv; Israel.\\
% ...
$^{166}$Department of Physics, University of Toronto, Toronto ON; Canada.\\
% ...

\end{flushleft}


\end{document}